\documentstyle[12pt,epsf]{article}
\catcode`\@=11 
\def\lsim{\mathrel{\mathpalette\@versim<}}
\def\@versim#1#2{\vcenter{\offinterlineskip
        \ialign{$\m@th#1\hfil##\hfil$\crcr#2\crcr\sim\crcr } }}
\catcode`\@=12 

\begin{document}
\baselineskip=19pt
\begin{titlepage}
\begin{flushright}
  KUNS-1678\\[-2mm]
  YITP-00-42\\[-2mm]
  hep-ph/0008069
\end{flushright}

\begin{center}
  \vspace*{1.5cm}
  
  {\large\bf Kaluza-Klein Mediated Supersymmetry Breaking}
  \vspace{8mm}
  
  Tatsuo~Kobayashi\footnote{E-mail address:
    kobayash@gauge.scphys.kyoto-u.ac.jp} and
  Koichi~Yoshioka\footnote{E-mail address:
    yoshioka@yukawa.kyoto-u.ac.jp}
  \vspace{3mm}
  
  $^*${\it Department of Physics, Kyoto University
    Kyoto 606-8502, Japan}\\
  $^\dagger${\it Yukawa Institute for Theoretical Physics, Kyoto
    University, Kyoto 606-8502, Japan}
  \vspace{1.5cm}
  
  \begin{abstract}
    We discuss a framework for communicating supersymmetry breaking to 
    the visible sector where the radius modulus, which determines the
    size of extra dimensions, has an auxiliary vacuum expectation
    value. The modulus couplings generate mass splitting in
    Kaluza-Klein supermultiplets and they act as messengers of
    supersymmetry breaking. The soft masses are expressed in terms of
    renormalization-group functions and the sparticle spectrum is
    determined by what kind of field propagates in the bulk. This
    framework also provides new possibilities for solving the
    supersymmetric flavor problem.
  \end{abstract}
\end{center}
\end{titlepage}


Low-energy supersymmetry (SUSY) provides an attractive candidate for
the fundamental theory beyond the standard model. However,
supersymmetry must be broken due to the absence of experimental
signatures below the electroweak scale. So far, various scenarios for
supersymmetry breaking have been proposed and for the scenarios to be
viable, several phenomenological constraints should be satisfied. For
example, we must have suppressions of new flavor-changing operators
which are generally introduced by superpartners of the standard-model
fields. This observation restricts possible forms of SUSY-breaking
terms~\cite{FCNC} and SUSY-breaking mechanisms at high-energy
scale. In particular, that may require there are no direct couplings
between the visible and SUSY-breaking sectors. In recently proposed
scenarios~\cite{RS,gauM} with extra dimensions, the SUSY-breaking
sector is spatially separated from the visible one literally, and the
flavor-changing operators are strongly suppressed. Another desirable
property of a SUSY-breaking scenario is that the sparticle spectrum is
distinctive of the scenario and detectable in future experiments. For
this issue, renormalization-group effects between the SUSY-breaking
scale and low-energy region would also be important. In this letter,
we discuss a new mediation mechanism of SUSY breaking in the model
with extra spatial dimensions.\footnote{For recent work for other
  SUSY-breaking mechanisms with extra dimensions, see
  Refs.~\cite{others}.}
As will seen below, in our scenario, the modulus which does not
directly couple to the visible sector has a non-vanishing $F$-term,
and moreover, characteristic spectra can be obtained depending on how
to construct four-dimensional low-energy effective theories.

We consider four-dimensional effective theories with additional
dimensions compactified at the scale $M_c$, which is smaller than the
cutoff $\Lambda$ of the effective theories. We assume for simplicity
that all the compactified dimensions have equal radius. The
compactification scale, i.e.\ the size of extra dimensions, is
determined by the vacuum expectation value of a radius modulus $T$. We
here assume that in the low-energy effective theories, the modulus
superfield has an auxiliary vacuum expectation value,
$\langle T\rangle=M_c+\theta^2F_T$. In this letter, we study the case
where the contribution of the radius modulus dominates SUSY
breaking. With other sources of SUSY breaking (nonzero $F$-terms),
there will exist additional effects on the soft SUSY-breaking
terms. When the standard-model fields propagate in the bulk, the
corresponding massive towers of Kaluza-Klein (KK) modes appear in the
four-dimensional effective theories. The mass couplings of chiral
supermultiplets for the $n$-th KK mode are given by the radius modulus
through the superpotential 
\begin{eqnarray}
  W &\sim& nT\,\Phi^{(n)}\,\bar\Phi^{(n)}.
\end{eqnarray}
Here $\bar\Phi$ is the SUSY-partner multiplet of $\Phi$ at each KK
level. This interaction induces mass splitting between the bosonic and
fermionic components of $\Phi$ and $\bar\Phi$, proportional to the
SUSY-breaking effect $F_T$. For KK vector multiplets, the mass
splitting in the multiplets comes from K\"ahler 
terms $\sim\int d\theta^2d\bar\theta^2 |nT|^2 V^{(n)}V^{(n)}$. The KK
excited modes thus work as messenger fields which transmit the
SUSY-breaking effect to the visible sector. In this scenario, the
visible sector consists of KK zero-modes and the fields confined on a
`3-brane'. It is noted that before Weyl-rescaling there are no
(non-derivative) direct couplings of the radius modulus $T$ to the
visible sector. (This issue is analyzed in minimal five-dimensional 
supergravity~\cite{LS}.) This is because $T$ can only couple to the
higher-dimensional components of the energy-momentum tensor and the
wave functions of massless fields do not depend on the
extra-dimensional coordinates. As a result, the soft SUSY-breaking
terms are generated at a quantum level in the four-dimensional
effective theory. That ensures the SUSY-breaking sector is hidden and
separated from our visible sector. 

We derive the expressions of soft SUSY-breaking terms. Let us first
consider the case where gauge fields propagate in the bulk whereas
other fields are confined on the 3-brane. The effective theory on the
brane contains the KK gauge and gaugino excited modes in addition to
other massless fields, KK zero-modes and boundary fields. The soft
terms for the latter massless fields can be extracted from
wave-function renormalization~\cite{extract}. Since there are a large
number of KK thresholds between $M_c$ and $\Lambda$, the dominant
SUSY-breaking effects come from the contribution of KK messengers. We
will neglect the usual logarithmic contributions in the following. In
this approximation, the one-loop renormalization-group equation of the
four-dimensional gauge coupling $\alpha$ is~\cite{DDG}
\begin{eqnarray}
  \frac{d\alpha}{dt} &=& \frac{b_{KK}}{2\pi}N(\mu)\alpha^2
  \label{alpha}
\end{eqnarray}
where $t=\ln\mu$ and $b_{KK}$ is the beta-function coefficient
determined by KK-mode contributions. The energy-dependent constant $N$
appears in summing up the leading-log contribution and indicates the
number of KK states propagating in the loop diagrams. It is
approximated by the volume of a sphere with radius $\Lambda/\mu$,
which is given by 
$N(\mu)\simeq\pi^{\delta/2}/\Gamma(1+\delta/2)(\Lambda/\mu)^\delta$ 
($\delta$ is the number of extra dimensions). The loop-expansion
parameter is roughly $N\alpha/4\pi$ and is assumed to be small. The
gaugino mass is extracted from an auxiliary component of the
$T$-dependent renormalized gauge coupling superfield. Integrating
eq.~(\ref{alpha}), we obtain the soft mass of the gaugino zero-mode at
the compactification scale,
\begin{eqnarray}
  M_g &=& b_{KK}N\frac{\alpha}{4\pi}\frac{F_T}{M_c}
  \label{gaugino}
\end{eqnarray}
where $N\equiv N(M_c)$. This result is exact at one loop and not
modified even when one considers other bulk interactions of KK states
like Yukawa couplings. As is well-known, the ratio $M_g/\alpha$ is
renormalization-group invariant and therefore eq.~(\ref{gaugino})
still holds below the compactification scale. Moreover, the ratio is
renormalization-group invariant under beta-functions due to KK
modes~\cite{kkmz}. Similar to gauge mediation, $M_g$ is proportional
to the gauge beta-function coefficient of messenger fields. Our
scenario is also similar to the moduli-dominated SUSY breaking in
superstring theory~\cite{string-T} where, for the vanishing
cosmological constant, soft terms are generated at loop level (string
corrections). In particular, the gauge kinetic function has
$T$-dependent threshold corrections due to KK modes. In this case, the
coefficient is written by the $N=2$ sector and the Green-Schwarz
coefficient due to the target-space duality
anomaly~\cite{T-anom}. Furthermore, it has been shown that the
$T$-dependent threshold corrections correspond to the power-law
behavior of the gauge coupling constant~(\ref{alpha}) in a certain
limit~\cite{GR}. On the other hand, in a scenario where bulk gauge
fields directly couple to SUSY-breaking sources, that gives rise to
tree-level soft masses for bulk zero-modes like the dilaton-dominated
SUSY breaking scenario.

Scalar soft masses are derived from wave-function renormalization of
four-dimensional chiral multiplets. For a chiral multiplet $\phi$, the
one-loop renormalization-group equation for the wave-function factor
$Z_\phi$ is
\begin{eqnarray}
  \frac{d\ln Z_\phi}{dt} &=& \frac{c_{KK}}{2\pi}N(\mu)\alpha
\end{eqnarray}
above the compactification scale. The coefficient $c_{KK}$ is the
quadratic Casimir invariant for the $\phi$ representation
($c_{KK}=4C_2(R_\phi)$). In a similar way to the gaugino mass, by
integrating the above renormalization-group equations, we obtain the
soft SUSY-breaking mass for the scalar component,
\begin{eqnarray}
  m_\phi^2 &=& \frac{\delta}{2}c_{KK}N\frac{\alpha}{4\pi}
  \left(\frac{F_T}{M_c}\right)^2,
  \label{gauge-m}
\end{eqnarray}
that is positive definite. We see that tachyonic modes generally do
not appear in contrast to, for example, anomaly
mediation~\cite{RS}. The soft mass terms are generated at one-loop
level and proportional to the number of extra dimensions,
$\delta$. These properties result from the fact that the
beta-functions of four-dimensional effective couplings explicitly
depend on the energy scale. The scalar trilinear term is also
calculated from the wave-function renormalization,
\begin{eqnarray}
  A_\phi &=& c_{KK}N\frac{\alpha}{4\pi}\frac{F_T}{M_c},
  \label{gauge-A}
\end{eqnarray}
and the trilinear coupling in the Lagrangian is given by the sum of
$A_\phi$'s of the fields forming a superpotential coupling 
(${\cal L}=-\sum A_\phi\phi\partial_\phi W(\phi)$). The trilinear
terms do not vanish at the boundary scale because of many KK
thresholds. Note that when a $\mu$-term generation mechanism is
specified, the corresponding holomorphic scalar soft mass, the
so-called $B$-term, is derived with the same form as
eq.~(\ref{gauge-A}).

In the present approximation, the SUSY-breaking effects are
communicated by the KK gauge multiplets that is flavor-blind. All
induced SUSY-breaking parameters become diagonal and degenerate in the 
flavor space, and consequently the flavor-changing neutral currents
(FCNC) are completely suppressed at the boundary scale. In a more
generic case, however, matter and Higgs fields also propagate in the
bulk. The corresponding KK modes and their couplings might give new
contributions to the FCNC processes. We will see below that even in
this case, there are new solutions to the FCNC problem, which are
peculiar to the models with extra dimensions.

With bulk matter and Higgs fields, there are additional contributions
to SUSY-breaking terms from the Yukawa couplings between the KK matter
and Higgs fields. That is, the radius modulus couplings, which induce
KK mass terms, give mass splittings in the KK multiplets at each
excited level, and the SUSY-breaking effect is transmitted to the
visible sector via the Yukawa couplings. In this case, there are two
points to notice. First, KK excited states generally belong to the
multiplets of $N=2$ or larger supersymmetry in the four-dimensional
effective theory. The radiative corrections from the KK multiplets
hence cancel out and SUSY-breaking terms are not generated. For
example, for bulk zero-modes, the gauge contributions disappear
($c_{KK}=0$) because of the hypermultiplet non-renormalization theorem
in unbroken $N=2$ theories. The second point is how to describe Yukawa
couplings among KK and boundary fields. We here assume that Yukawa
couplings between KK excited states arise from combinations of
trilinear and some higher-dimensional operators. On the other hand,
Yukawa couplings involving boundary fields are given by local terms in
higher-dimensional space~\cite{MP}. The soft SUSY-breaking terms
induced by Yukawa couplings have the expressions,
\begin{eqnarray}
  \Delta m^2 \;=\; -\frac{\delta}{2}a_{KK}N\frac{\alpha_y}{4\pi}
  \left(\frac{F_T}{M_c}\right)^2, \qquad
  \Delta A \;=\; -a_{KK}N\frac{\alpha_y}{4\pi}\frac{F_T}{M_c}.
  \label{yukawa}
\end{eqnarray}
Here $\alpha_y$ is the Yukawa coupling squared, $\alpha_y=y^2/4\pi$,
and the positive coefficient $a_{KK}$ denotes a multiplicity of the
Yukawa coupling in the one-loop anomalous dimension. The effects of
matter and Higgs KK messengers (\ref{yukawa}) are opposite in sign to
the gauge contributions (\ref{gauge-m}), (\ref{gauge-A}) and then
substantially change the sparticle spectrum. The gaugino mass is,
however, unchanged by the new interactions as noted before. It is
interesting that gathering together the above results, the KK-mediated
SUSY breaking are described in terms of the gauge beta-function
$\beta$ and the anomalous dimensions $\gamma_i$,
\begin{eqnarray}
  M_g &=& \frac{\beta}{2\alpha}\frac{F_T}{M_c}, \nonumber\\
  m_i^2 &=& -\frac{\delta}{2}\gamma_i\left(\frac{F_T}{M_c}\right)^2,
  \nonumber\\
  A_i &=& -\gamma_i\frac{F_T}{M_c}
  \label{formula}
\end{eqnarray}
for boundary fields and KK zero-modes. Scalar trilinear couplings have 
no new CP-violating phases. The renormalization-group functions
$\beta$ and $\gamma_i$ are determined by what types of multiplet live
in the bulk and on the four-dimensional brane. In other words,
sparticle masses measured in future experiments can restrict the
patterns of field configuration in the extra dimensions if the KK
mediation is dominant. Note that in our scenario, the messenger
interactions communicating SUSY breaking are related to the gauge and
Yukawa couplings in the standard model, which are experimentally
observed. Thus the particle and sparticle mass spectra correlate to
each other though supersymmetry is broken.

We now examine our results for the SUSY-breaking terms in several
examples. The first example is the model described in Ref.~\cite{DDG},
in which the standard model gauge and Higgs fields propagate in the
bulk while the chiral matter multiplets live on a 3-brane. We
calculate the soft terms with the formulas (\ref{formula}) in the
approximation with the three gauge couplings and the top Yukawa
coupling only. The beta-function coefficients are given by
$b_{KK}=(-6,-3,3/5)$ for $SU(3)$, $SU(2)$, and $U(1)$ respectively,
and we have the gaugino masses,
\begin{eqnarray}
  M_3 : M_2 : M_1 &=& 10\alpha_3 : 5\alpha_2 : \alpha_1.
  \label{gau1}
\end{eqnarray}
This ratio holds independently of the value of the compactification
scale $M_c$ and also remains unchanged under the renormalization-group
running below $M_c$. On the other hand, we have a variety of the
scalar soft masses. First, the Higgs soft masses vanish at the leading
order. This is because of the absence of radiative corrections from
the KK gauge messengers. One can hence get a natural electroweak
symmetry breaking without vacuum instability and fine-tuning of
parameters even in the case that running effects on the Higgs masses
are small, e.g.\ in the case of low-scale compactification. From this
point of view, it may be preferable that Higgs fields are of
higher-dimensional origin. Moreover, if the Higgs $\mu$-term is
induced by the Yukawa interaction with a gauge-singlet field like the
next-to-minimal supersymmetric standard model, the negative
contribution to the Higgs soft masses from the Yukawa coupling will
cause the symmetry breaking more naturally.

For the squark and slepton masses, we first find that this setup can
resolve the supersymmetric FCNC problem. Since the messenger Yukawa
couplings are small for the first two generations, the soft masses are
degenerate and the FCNC processes are suppressed. With
eqs.~(\ref{formula}), the sfermion masses are roughly on the order of
the gaugino masses but the top Yukawa effect reduces the stop
masses. Taking $M_c$ near below the GUT scale ($\simeq 10^{16}$ GeV)
and $\delta=1$ for example, we obtain
\begin{eqnarray}
  m_{{Q,u}_{1,2}}^2 \sim\, 1.7 m_{Q_3}^2 \,\sim\, 4 m_{u_3}^2 
  \,\sim\, m_d^2 \,\sim\, 2 m_L^2 \,\sim\, 2.5 m_e^2 \,\sim\, 2
  M_3^2
  \label{soft1}
\end{eqnarray}
at $M_c$. The relative ratios change through the renormalization-group
evolutions down to low energy. In this model, the lightest SUSY
particle is the bino (see eqs.~(\ref{gau1}) and (\ref{soft1})). The
bino mass becomes rather small compared to the gluino mass,
$M_3/M_1\sim 60$ at low energy, and that may conflict with the
experimental lower bound of $M_1$ and the naturalness problem. This
problem, however, can easily be avoided, for example, by adding bulk
fields charged under the $U(1)$ gauge symmetry. The detailed analyses
of spectrum, including the symmetry breaking issue etc., will be
discussed elsewhere. We finally comment that in this model, no-scale
supergravity-like boundary conditions 
($\sqrt{m^2}, A\ll M_g$)~\cite{noscale} may be viable if a small value
of our 3-brane tension would suppress the couplings between bulk and
boundary fields~\cite{BKNY}. In this case, the flavor-changing
operators are also suppressed.

The second example is the case that all the supersymmetric standard
model fields propagate in the bulk. If assuming the KK excited modes
consist in $N=4$ supermultiplets (i.e.\ a finite theory), they have no
effects on wave-function renormalization and the running behaviors of
couplings become logarithmic. The soft SUSY-breaking terms are then
determined by the beta-functions of the low-energy theory like the
case of anomaly mediation~\cite{RS,anomaly}. On the other hand, if the
KK excited modes belong to $N=2$ supermultiplets, the gaugino masses
are generated by nonzero gauge beta-functions while the scalar soft
terms vanish at one-loop level. We thus find a no-scale supergravity
boundary condition for the soft terms of the low-energy fields. In
this example, the gaugino masses become
\begin{eqnarray}
  M_3 : M_2 : M_1 &=& 6\alpha_3 : 9\alpha_2 : \frac{63}{5}\alpha_1.
\end{eqnarray}
The relation also holds at low energy. The other soft terms are
generated by the gaugino masses via renormalization-group running down
to low energy.

This example shows a new possibility for solving the supersymmetric
FCNC problem. As seen from (\ref{formula}), bulk zero-modes generally
have suppressed soft scalar masses compared to those of boundary
fields as well as gaugino masses. This is due to the restricted vertex
structure (non-renormalization property) of higher supersymmetry on
the KK excited states. Even when flavor-changing interactions
communicate SUSY breaking, their contributions to soft terms vanish
for KK zero-modes, and the FCNC processes involving these fields are
suppressed. Note that the situation is in contrast to the recently
proposed gaugino-mediation scenario~\cite{gauM}. There, bulk fields
couple to the SUSY-breaking source on a hidden brane and obtain soft
masses for zero-modes. As a result, matter multiplets are obliged to
exist only in the visible sector to suppress the dangerous FCNC
processes.

We finally comment on the gravitino mass scale. If we assume $F_T$ is
the only SUSY-breaking source, the gravitino mass $m_{3/2}$ is
estimated, by requiring a vanishing effective cosmological constant,
as $m_{3/2}\simeq F_T/M_{\rm pl}K(T)$. Here, $K$ is the function of
K\"ahler potential and $T$. Since we now consider $M_c<M_{\rm pl}$,
the gravitino becomes lighter than other SUSY particles and may be the
lightest one. That gives a cosmological bound 
for $m_{3/2}$ ($<O({\rm keV})$), and then for the magnitude of
$K$. (New inflation scenario could avoid this bound with a low
reheating temperature~\cite{new}.) For example, assuming the no-scale
type K\"ahler form, $K\sim T+T^*$, it leads an upper bound of the
compactification scale; 
$M_c\,\lsim\, 10^{-(2-3)} M_{\rm pl}\,\sim 10^{16}$ GeV.

In summary, we have shown that the non-vanishing $F$-term of the
radius modulus gives a new mediation mechanism for SUSY
breaking. There, KK excited states transmit the breaking effects to
the visible sector via the gauge and Yukawa couplings, which are
experimentally observed. The soft SUSY-breaking parameters are
expressed in terms of wave-function factors and the resulting spectrum
is shown to have rich structure depending on how the standard-model
fields extend in the bulk. The FCNC problem is solved with the
renormalization property of higher supersymmetry or a finite width of
3-brane, both of which can suppress the KK-mode couplings. Our SUSY
mediation mechanism also gives a framework of natural electroweak
symmetry breaking. It is noted that the spectrum is in contrast to the
case that there is a SUSY-breaking source somewhere in the extra
dimensions and bulk fields directly couple to it. In that case, the
bulk zero-modes obtain sizable soft scalar masses, which are generally
non-universal, but boundary fields have suppressed ones.

Note added: Similar ideas of using the $F$-component of the radius
modulus is discussed in Ref.~\cite{radion}.

\vspace*{5mm}
\subsection*{Acknowledgments}

The authors would like to thank to T.~Hirayama, K.~Kohri, J.~Kubo,
Y.~Nomura, S.~Sugimoto and M.B.~Tachibana for helpful discussions and
comments.



\vspace*{5mm}
\begin{thebibliography}{99}
\bibitem{FCNC}
  S.~Dimopoulos and H.~Georgi, {\sl Nucl.~Phys.} {\bf B193} (1981)
  150; J.~Ellis and D.V.~Nanopoulos, {\sl Phys.~Lett.} {\bf 110B}
  (1981) 44.
\bibitem{RS}
  L.~Randall and R.~Sundrum, {\sl Nucl.~Phys.} {\bf B557} (1999) 79.
\bibitem{gauM}
  D.E.~Kaplan, G.D.~Kribs and M.~Schmaltz, {\sl Phys.~Rev.} {\bf D62}
  (2000) 035010; Z.~Chacko, M.A.~Luty, A.E.~Nelson and E.~Ponton, 
  {\sl JHEP} {\bf 0001} (2000) 003.
\bibitem{others}
  I.~Antoniadis, {\sl Phys.~Lett.} {\bf 246B} (1990) 377;
  I.~Antoniadis, S.~Dimopoulos, A.~Pomarol and M.~Quiros, 
  {\sl Nucl.~Phys.} {\bf B544} (1999) 503; M.~Sakamoto, M.~Tachibana
  and K.~Takenaga, {\sl Phys.~Lett.} {\bf 458B} (1999) 231;
  N.~Arkani-Hamed, L.~Hall, D.~Smith and N.~Weiner, hep-ph/9911421.
\bibitem{LS}
  M.A.~Luty and R.~Sundrum, {\sl Phys.~Rev.} {\bf D62} (2000) 035008.
\bibitem{extract}
  G.F.~Giudice and R.~Rattazzi, {\sl Nucl.~Phys.} {\bf B511} (1998)
  25.
\bibitem{DDG}
  K.R.~Dienes, E.~Dudas and T.~Gherghetta, {\sl Phys.~Lett.} 
  {\bf 436B} (1998) 55; {\sl Nucl.~Phys.} {\bf B537} (1999) 47.
\bibitem{kkmz}
  T.~Kobayashi, J.~Kubo, M.~Mondragon and G.~Zoupanos, 
  {\sl Nucl.~Phys.} {\bf B550} (1999) 99.
\bibitem{string-T}
  L.E.Ib\'a\~nez and D.~L\"ust, {\sl Nucl.~Phys.} {\bf B382} (1992)
  305; B.~de~Carlos, J.A.~Casas and C.~Mu\~noz, {\sl Phys.~Lett.} 
  {\bf 299B} (1993) 234; A.~Brignole, L.E.Ib\'a\~nez and C.~Mu\~noz,
  {\sl Nucl.~Phys.} {\bf B422} (1994) 125; T.~Kobayashi, D.~Suematsu,
  K.~Yamada and Y.~Yamagishi, {\sl Phys.~Lett.} {\bf 348B} (1995) 402.
\bibitem{T-anom}
  J.-P.~Derandinger, S.~Ferrara, C.~Kounnas and F.~Zwirner, 
  {\sl Nucl.~Phys.} {\bf B372} (1992) 145.
\bibitem{GR}
  D.~Ghilencea and G.~Ross, {\sl Nucl.~Phys.} {\bf B569} (2000) 391.
\bibitem{MP}
  E.A.~Mirabelli and M.E.~Peskin, {\sl Phys.~Rev.} {\bf D58} (1998)
  065002.
\bibitem{noscale}
  For a review, A.B.~Lahanas and D.V.~Nanopoulos, {\sl Phys.~Rept.}
  {\bf 145} (1987) 1.
\bibitem{BKNY}
  M.~Bando, T.~Kugo, T.~Noguchi and K.~Yoshioka, 
  {\sl Phys.~Rev.~Lett.} {\bf 83} (1999) 3601.
\bibitem{anomaly}
  G.F.~Giudice, M.A.~Luty, H.~Murayama and R.~Rattazzi, {\sl JHEP}
  {\bf 9812} (1998) 027.
\bibitem{new}
  J.~Ellis, G.B.~Gelmini, J.L.~Lopez, D.V.~Nanopoulos and S.~Sarkar,
  {\sl Nucl.~Phys.} {\bf B373} (1992) 399; E.~Holtmann, M.~Kawasaki,
  K.~Kohri and T.~Moroi, {\sl Phys.~Rev.} {\bf D60} (1999) 023506.
\bibitem{radion}
  Z.~Chacko and M.A.~Luty, hep-ph/0008103.
\end{thebibliography}
\end{document}